\documentclass[aps,prl,floatfix,amsfonts,amssymb,amsmath,twocolumn,showpacs,preprintnumbers,nofootinbib,superscriptaddress]{revtex4}
\usepackage{graphicx}
\usepackage{amsmath,amsfonts,amssymb,amsthm}
\usepackage{fancyhdr}
\usepackage{mathrsfs}

\usepackage[latin1]{inputenc}
\usepackage[colorlinks]{hyperref}
\begin{document}

\title{Nonclassical photon pair production in a voltage-biased Josephson junction}

\author{Juha Lepp\"akangas}

\affiliation{Microtechnology and Nanoscience, MC2, Chalmers
University of Technology, SE-412 96 G\"oteborg, Sweden}

\author{G\"oran Johansson}

\affiliation{Microtechnology and Nanoscience, MC2, Chalmers
University of Technology, SE-412 96 G\"oteborg, Sweden}

\author{Michael Marthaler}

\affiliation{Institut f\"ur Theoretische Festk\"orperphysik
and DFG-Center for Functional Nanostructures (CFN), Karlsruhe Institute of Technology, D-76128 Karlsruhe, Germany}

\author{Mikael Fogelstr\"om}

\affiliation{Microtechnology and Nanoscience, MC2, Chalmers
University of Technology, SE-412 96 G\"oteborg, Sweden}


\begin{abstract}
We investigate electromagnetic radiation
emitted by a small voltage-biased Josephson junction connected to a superconducting transmission line. At frequencies below the well known emission peak at the Josephson frequency ($2eV/h$), extra radiation is triggered by quantum fluctuations in the transmission line. For weak tunneling couplings and typical ohmic transmission lines, the corresponding photon-flux spectrum is symmetric around half the Josephson frequency, indicating that the photons are predominately created in pairs.
By establishing an input-output formalism for the microwave field
in the transmission line,
we give further evidence for this nonclassical photon pair production, demonstrating that it violates the classical Cauchy-Schwarz inequality for two-mode flux cross correlations. In connection to recent experiments, we also consider a stepped transmission line, where resonances increase the signal-to-noise ratio. 
\end{abstract}

\maketitle


A voltage-biased Josephson junction (JJ) in series with a resistive environment
produces an oscillating supercurrent, known as the ac Josephson
effect~\cite{Josephson,Anderson,Likharev}.
This current creates electromagnetic (EM) radiation, usually in the microwave regime.
The voltage bias implies that the average voltage across the JJ is close to the total applied voltage.
The radiation has been studied intensively in
literature~\cite{Stephen1,Lee,DahmScalapino,Rogovin,Likharev,Levinson}, not least from a metrological perspective,
since its frequency $f$ is given by the applied voltage $V$, in the simplest form as $f=2eV/h$, where $e$ is the electron charge
and $h$ is Planck's constant.
For usual applications 
a classical treatment of the interplay is adequate.
Here, the microwave power spectrum has a peak at $f$ broadened by thermal fluctuations in the bias line.
At very low temperatures the quantum fluctuations of the resistor set a lower limit to the linewidth~\cite{Koch}, through a shot noise in the charge transport~\cite{IngoldEuro2002}.


In this Letter, we investigate the microwave field created by very small JJs,
when the charge transport takes place through an incoherent sequence of
independently tunneling Cooper pairs~\cite{Devoret,GirvinPRL1990,IngoldPE,Holst,Hofheinz}.
This was recently addressed experimentally~\cite{Hofheinz}, by simultaneous measurement of dc current and the power spectrum in the biasing transmission line.
It was found that the radiation consists either of a single photon or a pair of photons with the total energy $\sum_i hf_i=2eV$, corresponding to the  energy loss of a single tunneling Cooper pair.
The two-photon emission is triggered by quantum fluctuations
in the transmission line, and is of special interest, as this radiation is of nonclassical type.
Such production of nonclassically correlated photons has been
subject of an active study in systems taking advantage of non-linearities created by JJs~\cite{Yurke88,Yurke90,Johansson3,Eichler2011,Flurin2012,Beltran2008,Hakonen2011,Bergeal2010}.
It has a wide interest in the fields of quantum communication~\cite{cryptography} and metrology~\cite{spectroscopy1,spectroscopy2,spectroscopy3}.


\begin{figure}[b!]
\includegraphics[width=\linewidth]{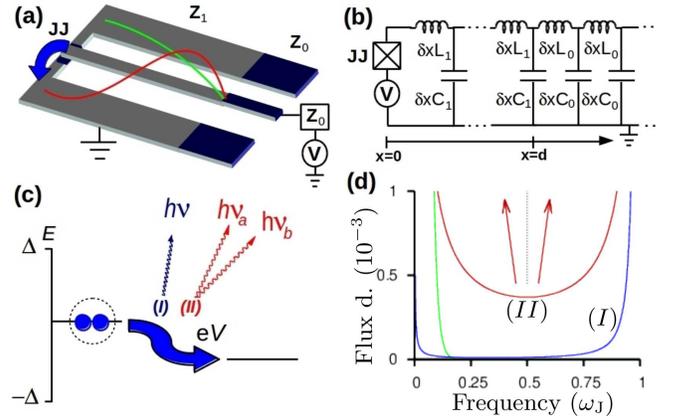}
\caption{(a)
A biased superconducting transmission line terminated by a small
Josephson junction (JJ). We consider an ohmic transmission line ($Z_0=Z_1$) and one with a step-like characteristic 
impedance ($Z_1\gg Z_0$), supporting the  visualized modes of a $\lambda/4$-resonator.
(b) The equivalent lumped-element model. The transmission line is characterized by its capacitance $C_{0}$ and inductance $L_0$ per unit length ($Z_i=\sqrt{L_i/C_i}$).
(c) The emitted radiation, either
single-photon ($I$) or two-photon emission ($II$), originates from the electrostatic energy released by a tunneling Cooper pair, $2eV$.
(d) The output photon-flux density due to thermal radiation $f_{\rm th}(\omega)$ (green), inelastic Cooper-pair tunneling $f_{\rm t}(\omega)$ [red, Eq.~(\ref{photon_spectrum})],
and as obtained from the classical treatment [blue, $k_{\rm B}T f_{\rm t}(\omega)/\hbar(\omega_{\rm J}-\omega)$].
The last two differ in emission at frequencies nearby
the half Josephson frequency $\omega_{\rm J}/2$, around which $f_{\rm t}(\omega)$
is symmetric. This indicates emission due to pair production of photons.
We use $k_{\rm B}T/2eV=0.017$ and $Z_0I_C/V=0.024$.
}
\label{fig1}
\end{figure}


We develop a theoretical framework for studying photon correlations. In particular, we establish
an input-output theory for the microwave field in a voltage-biased transmission line terminated by a JJ, see Fig.~\ref{fig1}a.
This provides a straightforward method to calculate any EM-field correlations in the output radiation
in the case of weak junction transparency.
Especially, by calculating photon-flux correlations in the output field, we show that the radiation is indeed of nonclassical type, as it violates a 
classical two-mode Cauchy-Schwarz inequality~\cite{Milburn,Silva,Li}. For optimal detection of this, we also consider a stepped transmission 
line~\cite{Hofheinz}, which leads to an increase in the signal-to-noise ratio by enhancing emission at specific frequencies.

The system we consider consists of a superconducting transmission line (TL)
terminated by a small Josephson junction (JJ) characterized by its capacitance $C_{\rm J}$ and critical current $I_{\rm c}$, see Fig.~\ref{fig1}a.
The line is dc-voltage biased,
with a voltage smaller than the superconducting gap, $eV<2\Delta$.
We focus on the EM radiation in the semi-infinite TL, which we describe by its magnetic flux in a discretized circuit model.
The TL is characterized by its capacitance $C_{0}$ and inductance $L_0$ per unit length, Fig.~\ref{fig1}b. We now treat the case of a homogenous TL  and discuss the stepped impedance case at the end of the paper.
In the continuum limit $\delta x\rightarrow 0$, the Heisenberg equations of motion for the flux field operator $\Phi(x,t)$ 
 has the form of a Klein-Gordon equation for a massless particle~\cite{Johansson2}, with the traveling-wave solution 
\begin{equation}\label{WaveSolution}
\begin{split}
&\Phi(x,t)=\sqrt{\frac{\hbar Z_{0}}{4\pi}}\int_0^\infty\frac{d\omega}{\sqrt{\omega}}\times \\
&\left[ a_{\rm  in}(\omega)e^{-i(k_\omega x+\omega t)}+a_{\rm out}(\omega)e^{-i(-k_\omega x+\omega t)}+{\rm H.c.} \right].
\end{split}
\end{equation}
Here $Z_0=\sqrt{L_0/C_0}$ is the characteristic impedance
and $k_{\omega}=\omega\sqrt{C_0L_0}$ the wave number.
The photon operators $a^{\dagger}_{\rm in/out}(\omega)$ $\left[a_{\rm in/out}(\omega)\right]$ create (annihilate) a photon of frequency $\omega$ moving leftwards (in) or rightwards (out) and satisfy the commutation relations
$
[a_{\rm in(out)}(\omega),a_{\rm in(out)}^{\dagger}(\omega')]=\delta(\omega-\omega')
$.
The left- and rightmoving part of the field are connected by the boundary condition imposed by JJ at $x=0$,
\begin{eqnarray}\label{RCJ}
C_{\rm J}\ddot\Phi(0,t)&+&\frac{1}{L_0}\frac{\partial\Phi(x,t)}{\partial x}\vert_{x=0}=\nonumber\\
&=&-I_{\rm c}\sin\left[2\pi\frac{\Phi(0,t)}{\Phi_0}-\omega_{\rm J} t\right].
\end{eqnarray}
Here $\Phi(x=0,t)$ is the magnetic flux across the junction, $\Phi_0=h/2e$ is the magnetic flux quantum, and $\omega_{\rm J}=2eV/\hbar$ is the Josephson frequency.

For a steady state one can solve for the output operators $a_{\rm out}(\omega)$ as a function of the input operators $a_{\rm in}(\omega)$.
We seek a solution in powers of the critical current $I_{\rm c}$ by multiplying the right-hand side of the boundary condition~(\ref{RCJ}) by $\xi$ and correspondingly write the solution for the outgoing wave in Eq.~(\ref{WaveSolution}) as $a_{\rm out}(\omega)=\sum_{n=0}^{\infty}\xi^n a_n(\omega)$. The input field is independent of $\xi$.
The zeroth-order solution (at $x=0$)
describes the phase shift given by reflection at the junction capacitance
\begin{equation}\label{ZerothOrder}
a_{0}(\omega)=[ \chi(\omega)/ \chi^*(\omega)]a_{\rm in}(\omega),
\end{equation}
where $\chi(\omega)=\sqrt{Z_0}/[1+i\omega/\omega_{\rm c}]$.
The cut-off frequency $\omega_{\rm c}=1/Z_0C_{\rm J}$ is given by the inverse
$RC$-time of the junction. It will be considered to be the highest
frequency scale in the problem.

The effect of Cooper-pair tunneling appears in the leading-order solution
\begin{eqnarray}\label{LeadingOrder}
a_1(\omega)=-I_{\rm c}\frac{i\chi(\omega)}{\sqrt{\hbar\omega \pi}} 
\int_{-\infty}^{\infty}dt\, e^{i\omega t}\sin[\phi_0(t)-\omega_{\rm J}t],
\end{eqnarray}
where
\begin{eqnarray}\label{ZerothPhaseFlucts}
\phi_0(t)=\frac{\sqrt{4\pi\hbar}}{\Phi_0}\int_{0}^{\infty}  \frac{d\omega}{\sqrt{\omega}}\chi(\omega) a_{\rm in}(\omega)e^{-i\omega t}+{\rm H.c.}
\end{eqnarray}
is the the zeroth order expression for the phase fluctuation operator at the junction $\phi(t)=(2\pi/\Phi_0)\Phi(0,t)$.

To second order in $I_{\rm c}$, we find
\begin{equation}\label{SecondOrder}
a_2(\omega)=I_{\rm c}^2\frac{i \chi(\omega)}{\sqrt{\hbar\omega \pi}} \int_{-\infty}^{\infty}dt e^{i\omega t}\left[\sin[\phi_0(t)-\omega_{\rm J}t],z(t)\right],
\end{equation}
where the operator 
\begin{equation}\label{SecondOrderSolution}
z(t)=\frac{i}{4e}\int_{-\infty}^{\infty}dt'\left[1+\frac{{\rm Sgn}(t-t')}{e^{\omega_c\vert t-t'\vert}-1} \right]\cos[ \phi_0(t')-\omega_{\rm J} t' ]
\end{equation}
is a solution to the equation
$\phi_1(t) =I_{\rm c}[z,\phi_0(t)]$, where $\phi_1(t)$ is the first-order result for the phase difference at the junction.
From Eqs.~(\ref{ZerothOrder}), (\ref{LeadingOrder}) and  (\ref{SecondOrder}) all correlation functions of the output field can be calculated, to second order in $I_{\rm c}$. This is the main theoretical result in this paper. As a consistency check, we have also verified that these perturbative expressions for the output operators satisfy the canonical commutation relations.

We can now straightforwardly calculate, e.g., the output photon-flux density, defined as~\cite{Loudon}
\begin{eqnarray}\label{flux}
f(\omega)=\int_0^{\infty} d\omega '\frac{1}{2\pi} \left\langle a^{\dagger}_{\rm out}(\omega)a_{\rm out}(\omega') \right\rangle,
\end{eqnarray}
up to second order in $I_{\rm c}$.
It can be  represented as a sum of two contributions: a part describing thermal radiation $f_{\rm th}(\omega)=(2\pi)^{-1}(\exp[\beta\hbar\omega]-1)^{-1}$
and a part  describing radiation originating in incoherent Cooper-pair tunneling  $f_{\rm t}(\omega)$, which dominates for $\hbar\omega\gg k_{\rm B}T$.
We now introduce the tunnel impedance~\cite{IngoldPE} $ {\rm Re}[Z_{\rm t}(\omega)] =|\chi(\omega)|^2=Z_0/\left[1+\left(\omega/\omega_c\right)^2\right]$.
The expression for the tunnel induced photon-flux density reads
\begin{equation}\label{OutputCorrelation}
f_{\rm t}(\omega)=\frac{I_{\rm c}^2  {\rm Re}[Z_{\rm t}(\omega)] }{2\omega}   \left[P(\hbar\omega_{\rm J}-\hbar\omega)+P(-\hbar\omega_{\rm J}-\hbar\omega)\right],
\end{equation}
where we have
\begin{eqnarray}\label{PEDefinition}
P(E)=\frac{1}{2\pi\hbar}\int_{-\infty}^{\infty}dte^{J(t)+i\frac{E}{\hbar}t},
\end{eqnarray}
and $J(t)=\left\langle [\phi_0(t)-\phi_0(0)]\phi_0(0) \right\rangle$ is the phase correlation function at the junction. The function $P(E)$ is the probability density for exchanging a total energy $E$ with the EM environment in a single tunneling event. Here the
term $P(\hbar\omega_{\rm d}-\hbar\omega)$ describes radiation coming from forward-direction Cooper-pair tunneling and the term
$P(-\hbar\omega_{\rm d}-\hbar\omega)$ from the Cooper-pair tunneling against the voltage bias, a process completely suppressed at temperatures $k_{\rm B}T \ll 2eV$.
The contribution $f_{\rm t}(\omega)$ agrees with Ref.~\onlinecite{Hofheinz}, obtained by applying the theory of inelastic Cooper-pair tunneling~\cite{IngoldPE}
to deduce the associated photon flux. The simultaneous dc current has then the form $I(V)=(\pi \hbar I_{\rm c}^2/4e)[P(2eV)-P(-2eV)]$.
We note that Eq.~(\ref{OutputCorrelation}) is valid also when the transmission line has resonances, a case treated at the end of this Letter.

The regime of validity for Eq.~(\ref{OutputCorrelation}) can be estimated by
comparing the phase fluctuations at the junction at different orders.
Since the right-hand-side of Eq.~(\ref{RCJ}) mixes all frequencies,
we demand the zeroth-order phase fluctuation spectrum to dominate at all $\omega$.
At frequencies $k_{\rm B}T/\hbar<\omega<\omega_{\rm J}-k_{\rm B}T/\hbar$
one obtains~\cite{LongPaper} a condition
$(2eI_{\rm c})^2{\rm Re}[Z_{\rm t}(\omega)]{\rm Re}[Z_{\rm t}(\omega_{\rm J}-\omega)]/\hbar^2\omega(\omega_{\rm J}-\omega)\ll 1$.
This implies for the low-ohmic transmission line $I_{\rm c}Z_0\ll V$.
At $\omega=0$ one gets $k_{\rm B}T/{\rm Re}[Z_{\rm t}(0)]\gg 4e I(V)$,
a comparison between Johnson-Nyquist current noise and the
transport current shot noise, also found in
Refs.~\cite{Likharev,DahmScalapino,Stephen1,Lee,Rogovin,Levinson,Koch,IngoldEuro2002}.
At $\omega=\omega_{\rm J}$ and for $Z_0=Z_1\ll R_Q$ we get $k_{\rm B}T\gg E_{\rm J}^2/\hbar\omega_{\rm J}$.
This can be translated
into a demand that the dephasing of the electromagnetic environment has to be much faster than the average Cooper-pair
tunneling rate.

We now focus on the {\em nonclassical origin} of the photon flux for frequencies $k_BT < \hbar\omega < \hbar\omega_d-k_BT$, where the simple TL environment allows for analytic expressions. At zero temperature, the classical result is $P(E)=\delta(E)$ [$J(t)=0$] and
the radiation power density becomes $I_{\rm c}^2{\rm Re}[Z_{\rm t}(\omega_{\rm J})]\delta(\hbar\omega_{\rm J}-\hbar\omega)/2$.
At finite temperatures the delta function broadens to a Lorentzian of width $\Gamma=4\pi k_{\rm B}T\rho$, where $\rho=4Z_0e^2/h$.
In a quantum treatment of the TL, vacuum fluctuations broadens $P(E)$ towards positive energies and more photons will be emitted at lower frequencies $\omega<\omega_{\rm d}$.
Then, for typical transmission line parameters, $\rho\ll 1$, the resulting
finite tail at lower frequencies has a simple expression,
\begin{equation}\label{photon_spectrum}
f_{\rm t}(\omega<\omega_{\rm J})= \frac{\rho I_{\rm c}^2Z_0 }{ \hbar\omega({\omega_{\rm J}-\omega}) }.
\end{equation}
This can be derived, e.g., using the long-time approximation~\cite{Devoret,Grabert,IngoldPRL},
$J(t)=-2\rho\left[ \ln(\omega_{\rm c} \vert t\vert)+\gamma + i\frac{\pi}{2}{\rm sign}(t) \right]$,
where $\gamma$ is the Euler constant.
The photon-flux density is symmetric around the half frequency $\omega_{\rm J}/2$
indicating that output radiation at these frequencies occurs through a pair production of photons symmetrically around $\omega_{\rm J}/2$, see Figs.~\ref{fig1}c-d.
This symmetry is a central result of this work, and also prevails to the case
of resonant transmission line.

A photon pair production in the microwave regime can be achieved also through
parametric effects in driven JJ-systems, through conversion of drive photons into the photon pairs~\cite{Yurke88,Yurke90, Johansson3,Eichler2011,Flurin2012,Beltran2008,Hakonen2011,Bergeal2010}.
In the present case, a static voltage bias is used and each photon pair is instead connected to a tunneling Cooper pair. We also note that the shape of the spectrum in Eq.~(\ref{photon_spectrum}) is inverted compared to the one obtained in the case of the dynamical Casimir effect (DCE)\cite{Johansson1} which is $\propto \omega(\omega_d-\omega)$. The difference arises since the Cooper-pair tunneling couples to phase fluctuations which are proportional to $1/\omega$, while for DCE the effective boundary inductance is modulated, which couples to current fluctuations proportional to $\omega$.

To get more proofs that photons are created in pairs, we evaluate the
second-order coherence $g^{(2)}(0)$\cite{Milburn} of the out-field, which gives the probability to measure two photons simultaneously normalized to the photon flux. To second order in $I_{\rm c}^2$ and for $\rho \ll 1$ we find~\cite{LongPaper}
\begin{equation}
g^{(2)}(0)=\frac{\langle a^{\dagger}_{\rm out}a^{\dagger}_{\rm out}a_{\rm out} a_{\rm out}\rangle}{\langle a^{\dagger}_{\rm out} a_{\rm out} \rangle^2} \approx 1+\left( \frac{2V}{R_Q I_{\rm c}} \right)^2.
\end{equation}
For low powers ($\propto I_{\rm c}^2$) this can be made arbitrary large, which indicates that (some part of) photons are indeed emitted in pairs. We can also verify,
that if the signal is filtered symmetrically around $\omega_{\rm J}/2$ the relative value of $g^{(2)}(0)$ increases.


To test whether the out-field also possesses {\em nonclassical correlations} we calculate the corresponding Cauchy-Schwarz (CS) inequality~\cite{Nonclassical1}.
In a two-mode case the inequality reads~\cite{Milburn,Li}
\begin{equation}\label{CSinequality}
\vert g_{ab}^{(2)}(0)\vert^2 \leq g_{aa}^{(2)}(0) g_{bb}^{(2)}(0),
\end{equation}
where the single-mode second-order coherence is defined as above
$ g_{ii}^{(2)}(0)=\langle a_i^{\dagger}a_i^{\dagger}a_ia_i\rangle/\langle a^{\dagger}_ia_i \rangle^2 $,
the photon-flux cross-correlator  
$ g_{ab}^{(2)}(0)=\langle a_a^{\dagger}a_a a_b^{\dagger} a_b\rangle/\langle a^{\dagger}_aa_a \rangle \langle a^{\dagger}_ba_b \rangle $,
and the indices $a$ and $b$ refer to two different modes.
We consider two frequency-separated modes obtained by detecting the out-field in a small frequency range $\delta\omega$ around $\omega_a$ and $\omega_b$~\cite{Silva}.
On the left-hand side (LHS) we have the squared probability of simultaneously observing a photon at frequency $\omega_a$ and $\omega_b$, which for a classical field is bounded by the right hand side (RHS) product of the probabilities of simultaneous pair detection at the individual frequencies.
We note that the out-field does not possess any steady-state single-mode or two-mode squeezing, in which case nonclassicality could have been shown measuring correlation functions to second order in the amplitude instead of flux\cite{Yurke90,Eichler2011,Flurin2012,Beltran2008}. The phase fluctuations across the junction influence the field in much the same way as pump phase fluctuations in a parametric amplifier, making the amplitude correlations short-lived.

\begin{figure}[t]
\begin{center}
\includegraphics[width=\linewidth]{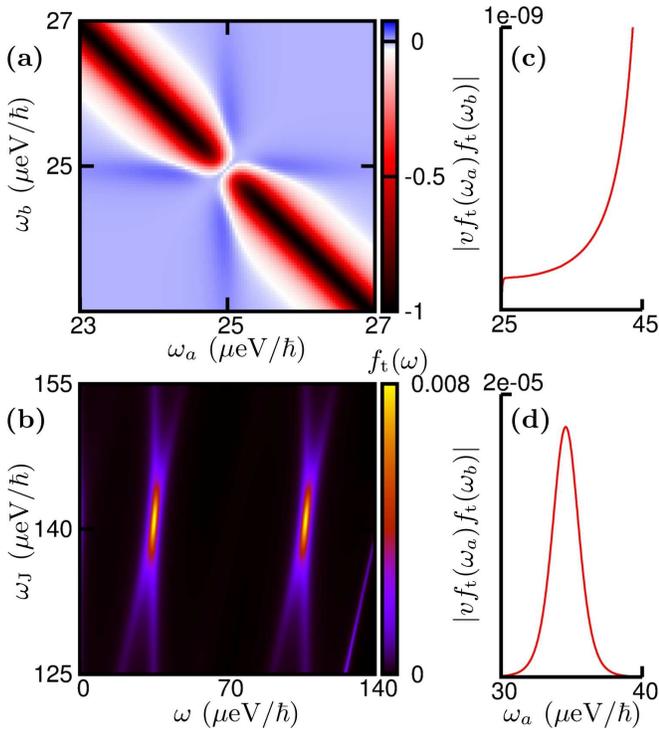}
\caption{(a) Violation of classical CS inequality for flux cross-correlations
between frequencies $\omega_a$ and $\omega_b$. 
Negative values are a sign of nonclassicality. The photon pair production is behind the strong violation nearby the diagonal
$\omega_a+\omega_b=\omega_{\rm J}=50$~$\mu$eV/$\hbar$. We plot here
${\cal V}/\vert {\rm Max}\{ {\cal V} \} \vert$, where ${\cal V}={\rm Sgn}(v)\ln(1+\vert v\vert)$
and
$v/(\mu eV)^2=
 P\left[\hbar(\omega_{\rm J}-2\omega_a)\right]   P\left[\hbar(\omega_{\rm J}-2\omega_b)\right]-P\left[\hbar(\omega_{\rm J}-\omega_a-\omega_b)\right]^2$. 
  We use $Z_0=Z_1=50$~$\Omega$, $C_{\rm J}=10$~fF, $T=100$~mK.
(b) Photon-flux density $f_{\rm t}(\omega)$ as a function of Josephson
frequency $\omega_{\rm J}$
in the neighborhood of $\omega_{\rm J}=\omega_0+\omega_1\approx 140$~$\mu$eV/$\hbar$,
when $Z_1\rightarrow 10Z_0$ and $E_{\rm J}=(\hbar/2e)I_{\rm c}=5$~$\mu$eV.
Here, the photon pairs are mainly emitted at $\omega_0\approx 35$~$\mu $eV/$\hbar$ and at $\omega_1\approx 3\omega_0$.
(c)
The {\em measured} violation of the CS inequality $-vf_{\rm t}(\omega_a)f_{\rm t}(\omega_b)$ in the diagonal $\omega_{\rm J}=\omega_a+\omega_b$ of (a).
(d) The measured violation corresponding to (b) when $\omega_a+\omega_b=\omega_0+\omega_1=\omega_{\rm J}$. The resonance occurs when $\omega_a=\omega_0$.
}
\label{fig2}
\end{center}
\end{figure}

With the assumption $\omega_a,\omega_b\gg k_{\rm B} T$
and in the limit $\delta\omega\rightarrow 0$ we get
to second-order in $I_{\rm c}$ the equivalent CS inequality for the out-field
\begin{equation}\label{Test2}
\begin{split}
 P\left[\hbar(\omega_{\rm J}-\omega_a-\omega_b)\right]^2 & \leq\\
 P\left[\hbar(\omega_{\rm J}-2\omega_a)\right]  & P\left[\hbar(\omega_{\rm J}-2\omega_b)\right].
\end{split}
\end{equation}
We notice that the nonclassicality of the field is determined by the $P(E)$-function only.
Now, at very low temperatures the probability to absorb energy from the EM environment goes to zero, i.e., $P(E<0)\approx 0$.
This implies that the RHS of Eq.~(\ref{Test2}) is close to zero for either $\omega_a>\omega_{\rm J}/2$ or $\omega_b>\omega_{\rm J}/2$. The LHS does not go to zero for $\omega_a+\omega_b \leq \omega_{\rm J}$, giving a large regime with possible nonclassical correlations. For $\omega_a=\omega_b$ (degenerate parametric down conversion) both sides are equal and the CS inequality cannot be violated.
This is the leading-order ($I_{\rm c}^2$) result, when all correlated photons originate from a single-Cooper-pair tunneling event.
The violation of inequality in Eq.~(\ref{Test2}) is visualized in Fig.~\ref{fig2}~a.



We estimate the photon-flux density in a typical ohmic transmission-line setup to be of the order $f(\omega_{\rm J}/2)\sim 10^{-4}$ photons per second per bandwidth, which should be readily detectable with a state-of-the art experimental setup~\cite{Hofheinz,Johansson3}. 
However, in the measurements of second order flux correlation functions the signal-to-noise ratio decreases substantially.
Thus, to verify the nonclassicality of the field it is favorable to use parameters maximizing the product of the photon flux at the detection frequencies $\omega_a$ and $\omega_b$, i.e., maximizing
$f_{\rm t}(\omega_a)f_{\rm t}(\omega_b)$.

To increase the photon flux, we consider resonantly shaping the $P(E)$-function using a simple step structure in the characteristic impedance~\cite{Hofheinz}, see Fig.~\ref{fig1}a.
For $Z_1\gg Z_0$ this provides modes at frequencies $\omega_n=(1+2n)\omega_0$.
Evaluating the output field at $x=d$, through similar theory as discussed before, we obtain that the main results Eqs.~(\ref{OutputCorrelation}) and (\ref{Test2}) prevail, after the change
\begin{equation}\label{ResonanceStructure}
\chi(\omega)= \frac{2\sqrt{Z_1} e^{-ik^1_\omega d}\sqrt{\frac{Z_1}{Z_0}}} { {\cal C}^{*}(\omega) \left(1+\frac{Z_1}{Z_0} \right)e^{-2ik_\omega^1 d}+{\cal C}(\omega)\left( \frac{Z_1}{Z_0}-1 \right) },
\end{equation}
where ${\cal C}(\omega)=1-iZ_1C_{\rm J}\omega $ and $k^1_{\omega}=\omega\sqrt{C_1L_1}$.
This changes the zeroth order phase fluctuation operator in Eq.~(\ref{ZerothPhaseFlucts}), and thus $P(E)$ through Eq.~(\ref{PEDefinition}) and the tunnel impedance ${\rm Re}[Z_{\rm t}(\omega)]=|\chi(\omega)|^2$.

By aligning the detection frequencies with the first two modes $\omega_a=\omega_0$ and $\omega_b=\omega_1$ and driving at $\omega_{\rm J}=\omega_a+\omega_b$  the photon flux can be enhanced dramatically.
In Fig.~\ref{fig2}b we plot numerical results for the output photon flux density for a setup with $Z_1=10Z_0$,
nearby the optimal drive for two-photon emission.
The output flux is mostly confined into the two frequencies (of the modes), being even larger than the flux at the Josephson frequency $\omega_{\rm J}$.
In Fig.~\ref{fig2}c-d we then plot the violation of the CS inequality at $\omega_{\rm J}=\omega_a+\omega_b$ for the open-space configuration and the resonant setup, multiplied by the product of the photon flux at the two detection frequencies.

In conclusion, we have derived expressions for the output field operators in a one-dimensional transmission line terminated by a voltage biased Josephson junction, to second order in the critical current of the junction. Using this formalism, we have confirmed the expression for the photon flux density derived in Ref.~\cite{Hofheinz} using a different formalism. Furthermore, by calculating second order flux correlation functions we have established that the photons below the Josephson frequency are mainly emitted in pairs and that the field is indeed nonclassical. Finally, we discuss the possibilities to enhance the photon flux by creating resonances in the transmission line and thus facilitate experimental detection.

We acknowledge financial support from the Swedish Research Council and from the EU through the projects ERC and the PROMISCE.

\end{document}